\documentclass[twocolumn]{aastex63}
\def \colw {0.5 \textwidth} 

\usepackage[english]{babel}
\usepackage[utf8x]{inputenc}
\usepackage[T1]{fontenc}
\usepackage{textgreek}

\usepackage{amsmath}
\usepackage{amssymb}
\usepackage{graphicx}
\usepackage{commath}
\usepackage{multirow}
\usepackage{color}
\usepackage{aecompl}
\usepackage{booktabs}
\usepackage[colorinlistoftodos]{todonotes}

\usepackage{tikz}

\newcommand{\I}{\protect\scriptsize I \normalsize $\!\!$}
\newcommand{\HI}{\mbox{\rm H\,\I}\: }
\newcommand{\HIit}{\mbox{\rm \textit{H}\,\scriptsize \textit{I} \normalsize $\!\!$}\: }
\newcommand{\HIns}{\mbox{\rm H\,\I}\,}

\shorttitle{OH Megamasers in \HI Surveys} 
\shortauthors{H. Roberts, J. Darling, A. J. Baker}

\begin{document}

\title{OH Megamasers in \HI Surveys:  Forecasts and a Machine Learning Approach to Separating Disks from Mergers}

\author[0000-0003-0046-9848]{Hayley Roberts}
\affil{Center for Astrophysics and Space Astronomy, 
Department of Astrophysical and Planetary Science, 
University of Colorado, 
389 UCB, Boulder, CO 80309-0389}

\author[0000-0003-2511-2060]{Jeremy Darling}
\affil{Center for Astrophysics and Space Astronomy, 
Department of Astrophysical and Planetary Science, 
University of Colorado, 
389 UCB, Boulder, CO 80309-0389}

\author[0000-0002-7892-396X]{Andrew J. Baker}
\affil{Department of Physics and Astronomy, 
Rutgers, The State University of New Jersey, 
136 Frelinghuysen Road, Piscataway, NJ 08854-8019}

\correspondingauthor{Hayley Roberts}
\email{hayley.roberts@colorado.edu}

\begin{abstract}
OH megamasers (OHMs) are rare, luminous masers found in gas-rich major galaxy mergers. In untargeted neutral hydrogen (\HIns) emission-line surveys, spectroscopic redshifts are necessary to differentiate the $\lambda_\text{rest}=18$ cm masing lines produced by OHMs from \HI 21 cm lines. Next generation \HI surveys will detect an unprecedented number of galaxies, most of which will not have spectroscopic redshifts. We present predictions for the numbers of OHMs that will be detected and the potential ``contamination'' they will impose on \HI surveys. We examine Looking at the Distant Universe with the MeerKAT Array (LADUMA), a single-pointing deep-field survey reaching redshift $z_\mathrm{HI}=1.45$, as well as potential future surveys with the Square Kilometre Array (SKA) that would observe large portions of the sky out to redshift $z_\mathrm{HI}=1.37$. We predict that LADUMA will potentially double the number of known OHMs, creating an expected contamination of 1.0\% of the survey's \HI sample. Future SKA \HI surveys are expected to see up to 7.2\% OH contamination. 
To mitigate this contamination, we present methods to distinguish \HI and OHM host populations without spectroscopic redshifts using near- to mid-IR photometry and a k-Nearest Neighbors algorithm. Using our methods, nearly 99\% of OHMs out to redshift $z_\mathrm{OH} \sim 1.0$ can be correctly identified. At redshifts out to $z_\mathrm{OH}\sim2.0$, 97\% of OHMs can be identified. The discovery of these high-redshift OHMs will be valuable for understanding the connection between extreme star formation and galaxy evolution.
\newline
\end{abstract}

\section{Introduction} \label{sec:intro}
OH megamasers (OHMs) are luminous 18 cm masers found in (ultra-)luminous infrared galaxies ([U]LIRGs) produced predominantly by major galaxy mergers. The dominant masing lines occur at 1667 and 1665 MHz in the ground state of OH. These masing lines have isotropic luminosities between $10^1-10^4 L_\odot$ and line widths ranging from 10 to 1000 km s$^{-1}$ due to Doppler broadening \citep{Darling2005}.  This rare phenomenon has only been discovered in roughly 110 galaxies out to redshift $z=0.264$ \citep{Darling2002a}.

OHMs are associated with high molecular gas density, $n(\text{H}_2)\sim 10^4 \text{ cm}^{-3}$, and strong far-IR radiation, making them markers for some of the most extreme star formation observed in our local universe \citep{Darling2007, Lockett2008}. These galaxies can provide significant information about how extreme star formation relates to galaxy evolution, particularly once discovered at higher redshifts. As products of gas-rich major mergers, OHMs can also provide an independent measure of the galaxy merger rate at a specific evolutionary stage. Further, the masing lines present in OHMs can be utilized as Zeeman magnetometers, providing in-situ measurements of magnetic fields in nearby galaxies \citep{Robishaw2008, McBride2014}. OHMs are useful astronomical tools for understanding many aspects of galaxy evolution.

However, in untargeted emission-line surveys for neutral hydrogen (\HIns), an OH emission line at $z_\mathrm{OH}$ can ``spoof'' a 21 cm \HI line at $z_\mathrm{HI}$ if $\nu_\mathrm{HI}/(1+z_\mathrm{HI}) = \nu_\mathrm{OH}/(1+z_\mathrm{OH})$ \citep{Briggs1997} where $\nu_\mathrm{HI} = 1420.4$ MHz and $\nu_\mathrm{OH} = 1667.4$ MHz. For example, restframe \HI corresponds to OH at a redshift of $z_\mathrm{OH} = 0.174$, while \HI at redshift $z_\mathrm{HI}=0.1$ would correspond to OH at a redshift of $z_\mathrm{OH}= 0.291$. These two lines have similar linewidths in their respective environments: \HI in spiral galaxies and OH in major galaxy mergers. Distinguishing between these lines often requires knowledge of a galaxy's spectroscopic redshift to determine the rest wavelength for an observed emission line (e.g., \citealt{Hess2021}). For galaxies that do not have spectroscopic redshifts, disentangling \HI from OH is not straightforward. This ambiguity particularly becomes a problem for high-redshift untargeted line surveys, which will not have prior spectroscopic redshifts for many of their detected galaxies. 

Despite the versatility and capacity of OHMs to serve as tools for studying galaxy evolution, line confusion in \HI surveys can serve as a source of contamination for the survey's main goals. Nonetheless, next-generation \HI surveys, such as those with the Square Kilometre Array (SKA) and its precursors, will be able to detect many new OHMs at redshifts never before reached. Finding these unique galaxies in the pool of \HI disk detections will enable an exciting new era in OHM science.

LADUMA (Looking at the Distant Universe with MeerKAT Array; \citealt{Blyth2016a}) is a survey with the MeerKAT radio interferometer, a precursor instrument for the SKA, that will be susceptible to OH/\HI confusion. LADUMA will be the deepest neutral hydrogen survey to date and is expected to detect \HI out to redshifts  $z_\mathrm{HI} =  1.45$. LADUMA's main science goals are related to studying the neutral atomic gas content of galaxies, meaning that OHM detections will contaminate its \HI samples. At low redshift, spectroscopic redshifts are generally known, so the contamination rate should be small. At greater distances where fewer spectroscopic redshifts are currently known and OHM prevalence increases due to the elevated merger rate, the contamination rate threatens to be higher. 

This paper presents predictions for the numbers of OH megamasers that will be detected by LADUMA in Section \ref{sec:NOH_in_LADUMA} and by other \HI surveys in Section \ref{sec:NOH_in_other_surveys}. We then present methods for distinguishing OH from \HI in LADUMA and other untargeted line surveys using near- to mid-IR photometry in Section \ref{sec:separating}. We discuss these results and their limitations in Section \ref{sec:discussion} and summarize our conclusions in Section \ref{sec:conclusions}.

Throughout this work, we assume a flat $\Lambda$CDM cosmology with $H_0 = 70$ km s$^{-1}$ Mpc$^{-1}$, $\Omega_m = 0.3$, and $\Omega_\Lambda = 0.7$.

\section{OH Megamasers in LADUMA} \label{sec:NOH_in_LADUMA}
In this section, we present our predictions for the number of OHMs that will be detected in LADUMA. A few inputs are needed to enable these predictions: an OH luminosity function, the volume of the survey, and the galaxy merger rate. The following subections will cover each of these. We then make use of this information to predict the number of OH megamasers and the resultant OHM contamination rate in other \HI surveys.

\subsection{The OH Luminosity Function} \label{sec:OHLF}
To create a prediction for the number of OHMs that will be found in the LADUMA survey, we need to integrate the OH luminosity function (OHLF) over the volume and luminosity limits of LADUMA. The OHLF presented in \cite{Darling2002} was constructed from OHMs detected by Arecibo and is valid for $2.2 < \log (L_\mathrm{OH}/L_\odot) < 3.8$ and $0.1<z_\mathrm{OH}<0.23$. An obvious source of uncertainty is the extrapolation of the OHLF to cover redshifts out to $z_\mathrm{OH}=1.876$ and luminosities as low as $1L_\odot$ for the LADUMA survey. 

The OHLF is defined as the number of OHMs with luminosity ($L_\mathrm{OH}$) per unit comoving volume ($\mathrm{Mpc}^3$) per logarithmic interval in $L_\mathrm{OH}$, and can be parameterized as
\begin{equation}
\Phi(L_\mathrm{OH}) = b\, L_\mathrm{OH}^a.
\end{equation}

The values of $a$ and $b$ presented in \cite{Darling2002} were determined using an error-weighted least-squares fit. We use a Markov chain Monte Carlo (MCMC) method to refit the OHLF using the data presented in \cite{Darling2002} to account for the correlations between fit parameters. We use $emcee$, a Python package for implementing MCMC \citep{emcee}. The refit OHLF is
\begin{multline} \label{eq:OHLF_me}
\Phi(L_\mathrm{OH}) = (3.17\pm0.58)\times 10^{-6} \ (L_\mathrm{OH}/L_\odot)^{-0.50 \pm 0.13} 
\\ \text{Mpc}^{-3}\ \text{dex}^{-1} .
\end{multline}
The samples from the MCMC allow us to create a representative sample of possible OHLF parameters for predictions of the number of OH megamasers that will be observed. This is the OHLF that will be used throughout this work. We later compare how the difference between the two OHLFs changes the number of OHMs predicted to be detected.

\subsection{Volume of LADUMA Field} \label{sec:volume}
LADUMA will be observing an area encompassing the Extended Chandra Deep Field South (E-CDFS), covering 0.9 deg$^2$ at $z=0$. The field of view of MeerKAT will increase with redshift due to the array's larger primary beam at lower frequencies. The total volume of the field can be calculated starting with the equation for comoving volume (equation 28 from \citealt{Hogg1999}) shown in equation (\ref{eq:hogg}), where $E(z) = \sqrt{\Omega_M (1+z)^3+\Omega_\Lambda}$, $D_H = c/H_0$, and $D_A$ is the angular diameter distance:
\begin{equation} \label{eq:hogg}
dV_c = D_H \frac{(1+z)^2 D_A^2}{E(z)} \ d\Omega \ dz.
\end{equation}
To account for an increasing field of view at higher redshifts, we write $d\Omega$ in equation (\ref{eq:hogg}) in terms of the redshift-dependent field of view of a telescope with primary beam diameter $1.22\, c\, (1+z) /( \nu_0\, D)$, where $D$ is the diameter of a single dish (for MeerKAT, $D = 13.5\ \mathrm{m}$) and $\nu_0$ is the rest frequency of the line being observed. This diameter yields a redshift-dependent solid angle:
\begin{equation}\label{eq:SA}
\Omega = \pi\Bigg[\frac{1.22}{2}  \frac{c \ (1+z)}{\nu_0\ D}\Bigg]^2.
\end{equation} 
Integrating equation (\ref{eq:hogg}) over $d\Omega$ and substituting in equation (\ref{eq:SA}) yields the differential volume for the LADUMA field:
\begin{equation}\label{eq:vol}
dV = \pi \bigg(\frac{1.22}{2}\bigg)^2 \bigg( \frac{D_A}{\nu_0 \ D}\bigg)^2  \frac{D_H\ (1+z)^4  }{\sqrt{\Omega_m (1+z)^3 + \Omega_\Lambda}} \ dz .
\end{equation}
Integrating equation (\ref{eq:vol}) over the redshift limits of the LADUMA survey for detecting OH ($z_\mathrm{OH}=0.174-1.876$), we obtain a total volume of $V=0.045 \text{ Gpc}^3$.

\subsection{Sensitivity of LADUMA}

\begin{figure}[t]
\centering
\includegraphics[width=\colw]{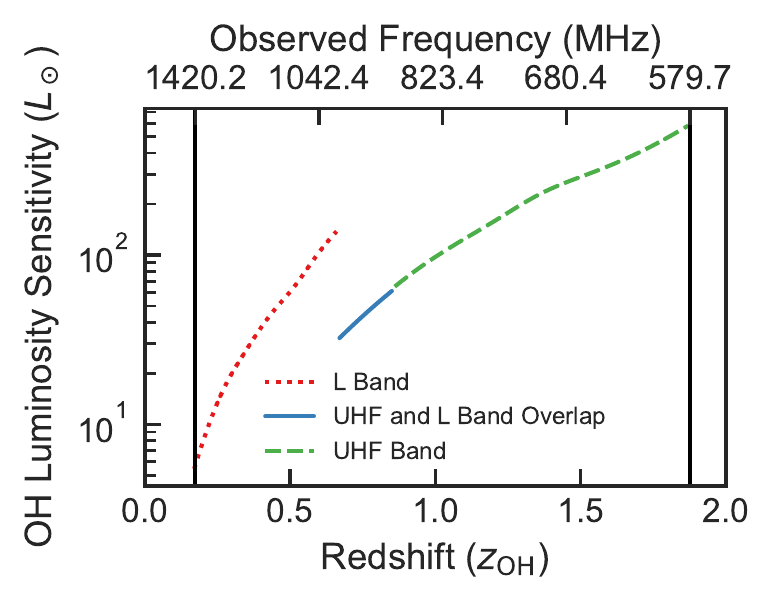} 
\caption{\label{fig:OHlum} Isotropic 5$\sigma$ luminosity sensitivity $L_\mathrm{OH}$ as a function of redshift and observed frequency. The vertical black lines show the low- and high-redshift limits of OH detection for LADUMA. The three different curves come from the two bands LADUMA will use, UHF and L bands, and the frequency range where they overlap. } 
\end{figure}
 
The luminosity sensitivity for integrating the OHLF changes with redshift and with MeerKAT band (either the UHF or L band). The sensitivity is calculated using equation (\ref{eq:sensitivity}), where $\Delta S_\nu$ is the interferometric sensitivity, SEFD is the system equivalent flux density as reported by MeerKAT\footnote{MeerKAT SEFD values can be accessed via the \href{https://skaafrica.atlassian.net/servicedesk/customer/portal/1/topic/bc9d6ad2-8321-4e13-a97a-d19d6d019a1c/article/277315585}{SARAO MeerKAT specifications page}.}, $\eta_\mathrm{corr}$ is the correlator efficiency, $N_\mathrm{ant}$ is the number of antennas, $N_\mathrm{pol}$ is the number of polarizations observed, $\Delta \nu$ is the bandwidth, and $\Delta t$ is the integration time:

\begin{equation} \label{eq:sensitivity}
\Delta S_\nu = \frac{\text{SEFD}}{\eta_\mathrm{corr}\sqrt{N_\mathrm{ant}(N_\mathrm{ant}-1)N_\mathrm{pol}\Delta\nu\Delta t}}.
\end{equation}
We assume that $\eta_\mathrm{corr} = 1$, all $N_\mathrm{ant} = 64$ antennas will be operating, and $N_\mathrm{pol}=2$. The assumed integration times are 333 hours for the L band and 3091 hours for the UHF band, with 3424 hours for the frequency range where the receivers overlap. We also assume a velocity width of $\Delta V =$ 150 km s$^{-1}$, the width of the average OH line at $z=0$, converted to hertz using $\Delta \nu/(\nu_\mathrm{OH}/(1+z))=\Delta V/c$. As LADUMA will be applying Briggs weighting to obtain a well-behaved synthesized beam, an extra noise penalty of 1.45 is included in sensitivity calculations. We also include the effects of primary beam attenuation away from the phase center, assuming two-dimensional Gaussian beams with half-power points defined as $0.5 \times 1.22 \, \lambda/D$. 

Using the sensitivity at each redshift, we calculate the $5\sigma$ luminosity limit as $L_\mathrm{min} = 4\pi D_L^2 \, 5\, \Delta S_\nu\, \Delta\nu $ where $D_L$ is the luminosity distance. Figure \ref{fig:OHlum} shows how the luminosity sensitivity changes with redshift for LADUMA. 

\subsection{Dependence on the Major Merger Rate} \label{sec:mergerrate}
An important element in the calculation of the number of OHMs is the galaxy merger rate, since OHMs arise in merging galaxies. This consideration leads us to introduce a factor of $(1+z)^\gamma$ in equation (\ref{eq:N}), where $\gamma$ is determined by the galaxy merger rate (in the sense of merging events per comoving volume) and will be referred to as the \textit{merger rate evolution coefficient}. Previous studies, based on both observations and simulations, have used different conventions for defining and parameterizing a ``merger rate,'' and drawn different conclusions that depend on type of merger, redshift, mass, and many other factors (e.g., \citealt{Lotz2011, Rodriguez-Gomez2015, Mundy2017, Mantha2018, Duncan2019, OLeary2020}, and references therein). The relevant merger rate for OHMs is one corresponding to gas-rich major mergers, whose evolution remains poorly constrained. Therefore, we select an intermediate estimate for this merger rate evolution coefficient of $\gamma \sim 2.2$ \citep{Rodriguez-Gomez2015, Mundy2017}, which will be assumed when not stated otherwise. In Section \ref{sec:finalN}, we present OHM calculations for a conservatively large range of possible $\gamma$ values, $0.0\leq \gamma\leq 3.0$, which is slightly larger than the range suggested by recent studies ($0.5\lesssim \gamma  \lesssim2.8$) \citep{Mundy2017,Ferreira2020}.

OHMs present a unique and independent way to measure the galaxy merger rate. Once we have a secure sample of OHMs, we will be able to provide an estimated $\gamma$ for gas-rich major mergers. As more \HI surveys take data, this method will be a robust way for tracing the cosmic history of major mergers.

\subsection{MCMC Calculation of N(OH)} \label{sec:finalN}
Using the MCMC fit samples for the OHLF discussed in Section \ref{sec:OHLF}, we can integrate over volume, luminosity, and merger rate to get a prediction of the number of OH megamasers to be detected at the 5$\sigma$ level in LADUMA as shown in equation (\ref{eq:N}):

\begin{equation} \label{eq:N}
N = \int_{L_\mathrm{min}}^{L_\mathrm{max}} \int_{0}^{V_\mathrm{total}} \Phi(L_\mathrm{OH}) \ (1+z)^\gamma \ dV \ d\log L_\mathrm{OH}.
\end{equation}

\begin{figure}[t]
\centering
\includegraphics[width =\colw]{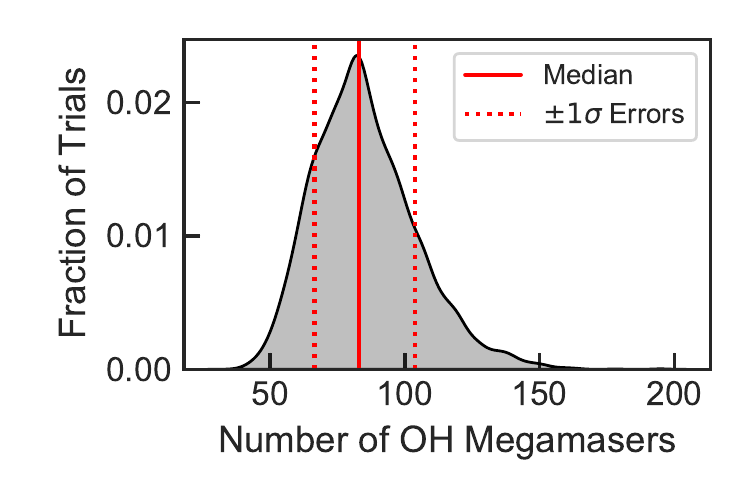} 
\caption{\label{fig:NOH} Kernel density estimation of the predicted number of OHMs to be found by LADUMA, assuming $\gamma=2.2$, using 10,000 samples from the MCMC fit to the OHLF parameters. The expected number of OHM detections is $83^{+21}_{-17}$.} 
\end{figure}

Figure \ref{fig:NOH} shows a distribution of possible values for the number of OHMs using 10,000 samples from the MCMC fit to the OHLF, assuming that $\gamma = 2.2$. The median value is 82.99, and the associated 16$^\text{th}$ and 84$^\text{th}$ percentiles are 66.49 and 103.89. The 16$^\text{th}$ and 84$^\text{th}$ percentiles are commonly associated with $-1\sigma$ and $+1\sigma$ limits (for a Gaussian distribution) respectively, and they will be referred to as such for the remainder of this paper. The mean of the distribution is $85.32$. The number of OH megamasers we expect to detect with LADUMA is therefore $83^{+21}_{-17}$. This total would nearly double the number of known OHMs.

\begin{figure}[t]
\centering
\includegraphics[width =\colw]{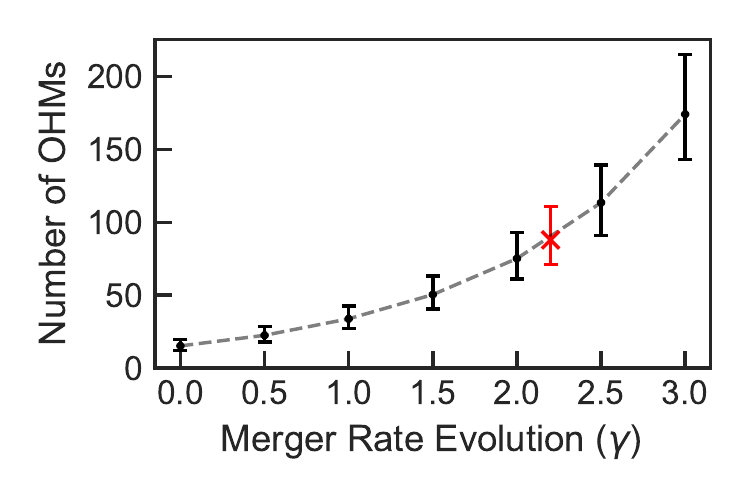} 
\caption{\label{fig:merger} Number of OHM detections in LADUMA versus merger rate evolution parameter $\gamma$. The fiducial $\gamma = 2.2$ is denoted by a red X marker.}
\end{figure}

As discussed in Section \ref{sec:mergerrate}, the exact value of the merger rate evolution parameter $\gamma$ is poorly constrained. Figure \ref{fig:merger} presents the above calculation for values of $\gamma$ ranging from 0.0 to 3.0. If $\gamma$ is assumed to be 0.0, and the merger rate does not increase with redshift, then it is expected that LADUMA would detect $15^{+5}_{-3}$ OHMs.

\begin{figure}[t]
\centering
\includegraphics[width =\colw]{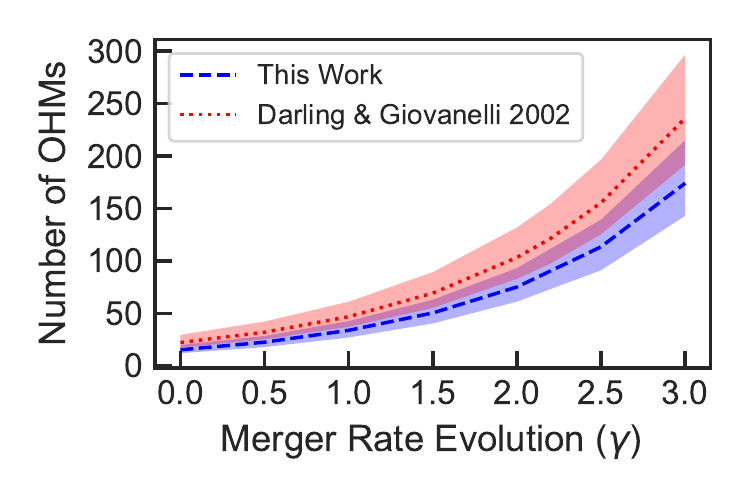} 
\caption{\label{fig:NOHComparisonDG} Comparison of OHM detection rate in LADUMA versus merger rate evolution coefficient for the OHLFs presented in this work and in \cite{Darling2002}. The dashed and dotted lines show the median values, and the shaded areas represent $\pm 1 \sigma$ uncertainties.}
\end{figure}

We can compare the impact of the OHLF presented in this work to the OHLF in \cite{Darling2002} by calculating the numbers of OHMs implied by the two. Figure \ref{fig:NOHComparisonDG} shows how the number of OHMs varies for the two OHLFs. When $\gamma = 2.2$, the MCMC approach adopted here implies a factor of 1.4 fewer detections. In general, the larger uncertainties from calculations with the OHLF in \cite{Darling2002} are due to the larger uncertainties in that paper's OHLF parameters. 

\section{OH Contamination in Other \HI Surveys} \label{sec:NOH_in_other_surveys}
In this section, we generalize the above calculations for \HI surveys underway or planned at other radio telescope arrays, particularly the Australian Square Kilometre Array Pathfinder (ASKAP), the Very Large Array (VLA), the APERture Tile In Focus (APERTIF) facility, and Phase I of the SKA. The VLA and ASKAP have \HI surveys underway or planned. At the VLA, the COSMOS \HI Large Extragalactic Survey (CHILES), a single-pointing survey, is currently analyzing data \citep{Fernandez2016}. WALLABY (Widefield ASKAP L-band Legacy All-sky blind surveY) is in the process of observing pilot fields with ASKAP alongside DINGO (Deep Investigation of Neutral Gas Origins; \citealt{Duffy2012}). DINGO will have two tiers, Deep and Ultra Deep. For our analysis, we consider only DINGO-Deep. APERTIF is also currently collecting data on the Westerbork Synthesis Radio Telescope (WSRT) and will execute multiple surveys at different depths. We consider the Medium Deep Survey (MDS) as described in the APERTIF Survey Plan. 
The SKA is a future telescope that will come in two phases, with the first (SKA1) covering $\sim$10\% of the total collecting area of the second (SKA2; \citealt{Abdalla2014}). Due to the uncertainties in the schedule for SKA2, we consider only possible mid-frequency SKA1 surveys for this analysis.

\begin{deluxetable}{lccc}[!t]
\tablecaption{\HI Survey Parameters for $N$(OH) Calculation\label{tab:HI_survey_info}}
\tablehead{\colhead{Survey} & \colhead{Redshift Range} & \colhead{Sky Area} & \colhead{$5\sigma$ Sensitivity} \\ 
\colhead{} & \colhead{($z_\mathrm{HI}$)} & \colhead{(deg$^2$)} & \colhead{(μJy)} } 

\startdata
LADUMA					 & 0.0--1.45   	& 0.90\tablenotemark{a} 	& 48\tablenotemark{b} \\
CHILES						 & 0.0--0.45  	& 0.32\tablenotemark{a}	& 350 \\
WALLABY					 & 0.0--0.26	& 27,500  							& 5,000 \\
DINGO-Deep				 & 0.0--0.26	& 150       							& 156   \\
APERTIF MDS			 & 0.0--0.26	& 450       							& 1250  \\
SKA1 Medium wide	 & 0.0--0.29	& 400       							& 247\tablenotemark{b}   \\
SKA1 Medium deep	 & 0.0--0.50 	& 20        							& 52\tablenotemark{b}    \\
SKA1 Deep				 & 0.35--1.37	& 1\tablenotemark{a}        	& 13\tablenotemark{b}    \\
SKA1 All-sky				 & 0.0--0.50 	& 20,000    						& 525\tablenotemark{b}   \\
\enddata
\tablenotetext{a}{Single-pointing survey comoving volume calculations include an expanding field of view at higher redshifts and lower observing frequencies.}
\tablenotetext{b}{LADUMA and SKA1 have published frequency-dependent sensitivities, which we employ for calculations in this paper. The values presented here are the mean sensitivities across the entire range of the observed frequencies.}
\tablerefs{CHILES \citep{Fernandez2013}, WALLABY \& DINGO-Deep \citep{Duffy2012}, APERTIF (\href{https://old.astron.nl/sites/astron.nl/files/cms/OTHER/ApertifSurveyPlanII.v2.2.pdf}{Apertif Survey Plan}), SKA1 \citep{Braun2019,Staveley-Smith2014}}
\end{deluxetable}

The calculation of numbers of OHMs for these \HI surveys uses equation (\ref{eq:N})  above. For each survey, we calculate $L_\mathrm{min}$ as a function of redshift using the reported sensitivity and we calculate the volume using redshift ranges and sky coverage. All assumptions made about the survey or telescope for these calculations are presented in Table \ref{tab:HI_survey_info}. The sensitivity column assumes a velocity width of 150 km s$^{-1}$ as done for the LADUMA calculations. Concepts for the SKA1 surveys come from \cite{Staveley-Smith2014}. Each of the three fiducial surveys (medium wide, medium deep, and deep) assumes a total observing time of 1,000 hours, while the all-sky commensal survey assumes an observing time of 10,000 hours. \cite{Staveley-Smith2014} note that angular resolutions finer than 10" are only accessible to the SKA for high column densities --- a limitation that especially applies to the SKA1 Deep survey, whose nominal angular resolution is 2". We have not modeled the effects of resolution on the detectability of OHMs, but we note that at high resolutions, there could be a bias in favor of detecting (more compact) OHMs relative to (more extended) \HI emitters.\footnote{This paper's predictions ignore the effects of radio frequency interference, which can vary for different sites and different array configurations. Here too, it may in practice be systematically easier to recover more compact OHMs than more extended \HI emitters in frequency ranges where RFI precludes the use of short-baseline data.}

For surveys featuring single pointings, we have calculated comoving volumes assuming that sky area increases $\propto (1 + z_{\rm H\,I})^2$ due to the increasing size of the primary beam at lower frequencies. This calculation applies to CHILES, LADUMA, and a hypothetical SKA1 Deep survey and is noted in Table \ref{tab:HI_survey_info}. For surveys covering larger sky areas through the use of multiple pointings across contiguous patches, sky area will be higher at $z_{\rm H\,I} > 0$ than at $z_{\rm H\,I} = 0$, but the change will be less dramatic because only the pointings that lie at the edges of the contiguous patches will contribute. Because this effect will be small in a fractional sense (smaller for larger sky areas), and will depend on the detailed distribution of patch sizes, we do not correct for it. We also choose to omit primary beam attenuation when predicting OHM contamination of other \HI surveys, since we cite values for the numbers of \HI detections that do not include this consideration \citep{Staveley-Smith2014}.

\begin{deluxetable}{lccc}[!th]
\tablecaption{Predicted 5$\sigma$ OHM and \HI Detections for Untargeted \HI Surveys\label{tab:NOH_surveys}}
\tablehead{\colhead{Survey} & \colhead{$N$(OH)} & \colhead{$N$(\HIns)} & \colhead{$N$(OH)/$N$(\HIns)}} 
\startdata
LADUMA                      & $8.3^{+2.1}_{-1.7}\times10^1$    & $8\times10^3$   & $1.03^{+0.26}_{-0.20}\%$  \\
CHILES                      & $5.5^{+1.1}_{-0.9}\times10^{-1}$ & $3\times10^2$   & $0.18^{+0.04}_{-0.03}\%$  \\
WALLABY                     & $8.9^{+2.5}_{-1.8}\times10^2$    & $6\times10^5$   & $0.15^{+0.04}_{-0.03}\%$  \\
DINGO-Deep                  & $7.5^{+2.3}_{-1.7}\times10^1$    & $5\times10^4$   & $0.15^{+0.04}_{-0.03}\%$  \\
APERTIF MDS        & $2.6^{+0.6}_{-0.5}\times10^2$    & $3\times10^5$   & $0.09^{+0.02}_{-0.01}\%$  \\
SKA1 Medium wide    & $2.5^{+0.7}_{-0.6}\times10^2$    & $3.4\times10^4$ & $0.73^{+0.21}_{-0.17}\%$  \\
SKA1 Medium deep    & $7.7^{+3.2}_{-2.3}\times10^1$    & $2.5\times10^4$ & $0.31^{+0.13}_{-0.09}\%$  \\
SKA1 Deep           & $1.9^{+0.6}_{-0.5}\times10^2$    & $2.6\times10^3$ & $7.20^{+2.32}_{-1.81}\%$  \\
SKA1 All-sky        & $4.1^{+1.0}_{-0.8}\times10^3$    & $5.5\times10^5$ & $0.75^{+0.18}_{-0.14}\%$  \\
\enddata
\tablecomments{$N$(OH) values assume merger rate evolution coefficient $\gamma = 2.2$.}
\end{deluxetable}

Table \ref{tab:NOH_surveys} presents the number of OHMs predicted to be detected in each survey for merger rate evolution coefficient $\gamma = 2.2$. Table \ref{tab:NOH_surveys} also presents the number of \HI sources each survey expects to detect. The contamination column is the ratio of OHM detections to \HI detections, which can be related to the fraction of an ``\HI sample'' that will actually be OHMs mistaken for \HI sources if spectroscopic redshifts are unavailable.

One noteworthy aspect from Table \ref{tab:NOH_surveys} is the much higher rate of contamination for LADUMA and the SKA1 Deep survey compared to the other surveys. These are distinctly different from the other \HI surveys due the fact that they extend to significantly higher redshift. We therefore infer that \HI detections dominate at low redshifts (i.e., $z_\mathrm{HI}\lesssim 1$) for all surveys. However, the OH detection density surpasses the \HI detection density at redshifts above $z_\mathrm{HI} \sim 1$ --- an effect that is pronounced for single-pointing surveys that have much larger relative fields of view at high versus low redshifts.


An earlier conclusion in the same vein was reached by \cite{Briggs1997}, who predicts that OHM contamination in \HI surveys will increase with redshift. We explore how that contamination depends on redshift using both LADUMA and an expanded version of the SKA1 Deep survey for comparison. The SKA1 Deep survey presented in \cite{Staveley-Smith2014} only covers a frequency range of 600--1050 MHz that corresponds to only a portion of the full SKA1 mid-band \citep{Braun2019}. For the purpose of exploring OH contamination versus redshift, we assume a deep survey that exploits the full range of the mid-band and therefore covers a frequency range of 600--1420 MHz. Equation (\ref{eq:N}) is used to estimate how the number of OHMs varies with redshift. To estimate the number of \HI detections per redshift interval, we use the following equation from \cite{Obreschkow2009}:

\begin{equation}
\frac{dN/dz}{\text{1 deg}^2} = 10^{c_1} \ z^{c_2} \ e^{-c_3 z},
\end{equation}
where the $c_i$ are parameters specific to each \HI survey. LADUMA's values are interpolated from \cite{Obreschkow2009} for each redshift using sensitivities calculated from equation (\ref{eq:sensitivity}) as the limiting integrated flux, and assuming linewidths of 100 km s$^{-1}$, allowing us to determine \HI detection rate versus redshift. We follow a similar method for the SKA1 Deep survey using sensitivity values presented in \cite{Braun2019}. We calculate how number density for OH and \HI varies with redshift, assuming redshift bins of $dz=0.01$. \cite{Obreschkow2009} note that $dN/dz$ will be $\geq 1\%$ underestimated for $z\leq z_c$ where $z_c$ depends on the limiting flux. For both LADUMA and SKA1 Deep, on average, $z_c\sim 0.1$. Therefore, the numbers of \HI detections are slightly underestimated for $z_\mathrm{HI}\leq 0.1$.

\begin{figure}[t]
\centering
\includegraphics[width =\colw]{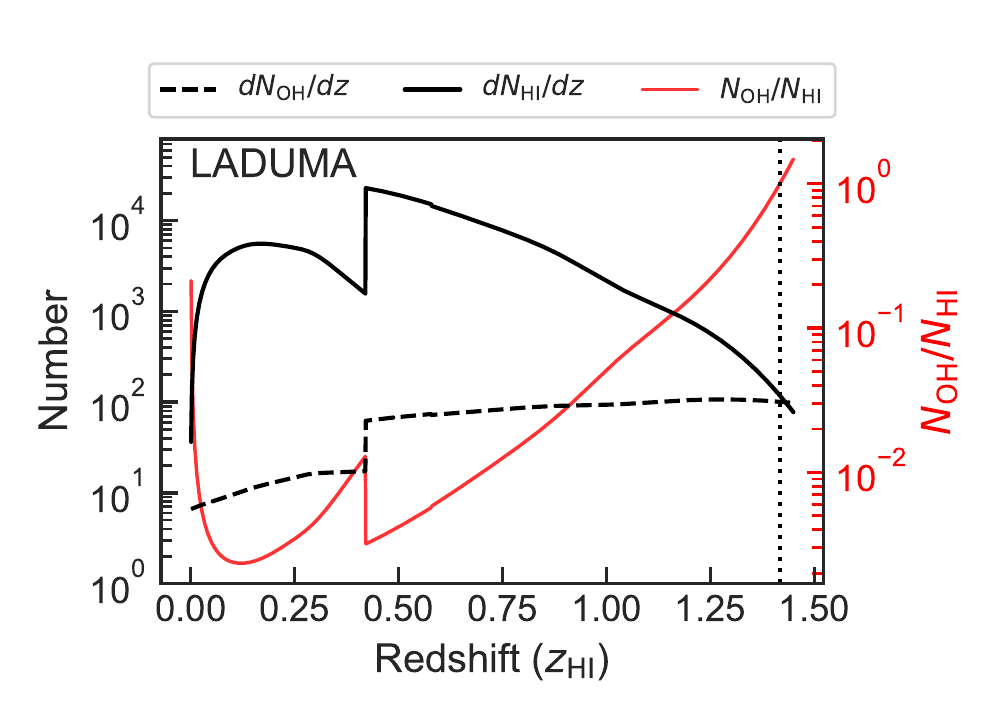} 
\includegraphics[width =\colw]{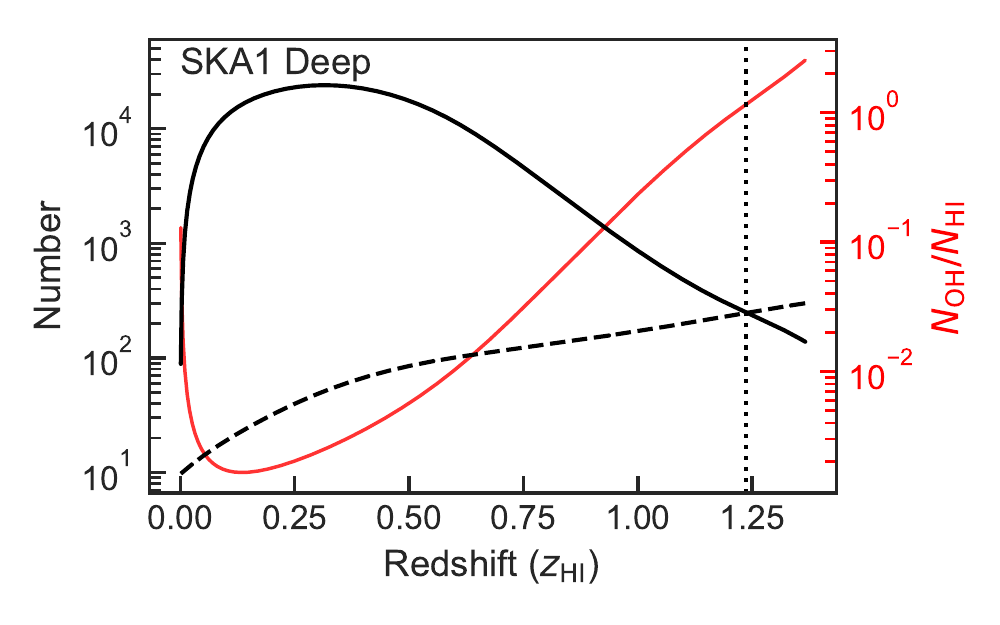} 
\caption{\label{fig:NOHvsNHI} Projected numbers of OHMs and \HI sources and the OHM fraction $(\mathrm{N}_\mathrm{OH}/\mathrm{N}_\mathrm{HI})$ vs redshift $z_\mathrm{HI}$ for LADUMA (above) and SKA1 Deep (below). Left axis shows numbers of objects (plotted in black); the dashed curve indicates how the number of OH detections changes with redshift, and the solid curve indicates the same for \HI detections. Right axis shows the ratio of OHM detections to \HI detections (plotted in red). The vertical dotted lines indicate the redshifts where the numbers of OH and \HI detections are equal. The discontinuities in the LADUMA curves originate from the overlap between the L and UHF bands. All calculations assume galaxy merger rate evolution parameter $\gamma=2.2$.}
\end{figure}

Figure \ref{fig:NOHvsNHI} shows how the detection rate of OH and \HI varies with redshift. \HI detections dominate at low redshift for both surveys. At higher redshifts, the OH detection rate and OH fraction grow significantly. This comparison also demonstrates how the OHM contamination rate depends on survey parameters, as discussed in Section \ref{sec:NOH_in_other_surveys}.  
For LADUMA, OHMs will not outnumber \HI source at any redshift probed by the survey (i.e., for any $z_\textrm{HI} \leq 1.45$); in comparison, SKA1 Deep's sensitivity as a function of frequency will yield a number of OH detections surpassing that of \HI detections for $z_\mathrm{HI}\geq 1.25$.

\cite{Briggs1997} first presented this issue of OH contamination in untargeted \HI surveys. Results from that paper demonstrated that by redshift z$_\mathrm{HI}\sim$ 1.0, OH detections would significantly outnumber \HI detections for a much shallower survey ($5\sigma = 5$ mJy) than LADUMA or SKA1 Deep. 
The predicted transition from an \HIns -dominated to an OH-dominated sample is qualitatively consistent with our findings, although the precise redshift at which this happens depends on survey parameters.  In general, the deeper a survey observes over a given frequency range, the more \HI emitters it will detect relative to OHMs.  LADUMA and SKA1 Deep are
both deeper than the hypothetical \cite{Briggs1997} survey, allowing them to detect more \HI sources at higher redshifts and thus to push out the projected redshift at which OH detections outnumber \HI detections.  The fact that this transition
occurs at a lower redshift ($z_{\rm H\,I} = 1.25$) for SKA1 than for LADUMA owes to the fact that LADUMA's sensitivity improves at higher redshifts (due to its distribution of observing time), in contrast to the SKA1 Deep sensitivity.

\newpage
\section{Identifying OH Megamasers in Untargeted \HI Surveys} \label{sec:separating}
Distinguishing an OH from an \HI line is currently only done using the optical spectroscopic redshift of an object to determine an observed line's rest wavelength. Next-generation \HI surveys will observe orders of magnitude more objects than previous surveys, as shown in Table \ref{tab:ML_results}, most of which will not have spectroscopic redshifts available. For that reason, we explore machine learning as a way to distinguish OH from \HI emission lines using ancillary data.

\subsection{Creating OH and \HIit Models for Distinguishing Populations}
\subsubsection{Fitting OHM Host Galaxy SEDs}
The limited number of OHMs creates serious limitations in understanding the OH population and how it differs from \HI hosts.
We therefore fit the spectral energy distributions (SEDs) of 111 OHMs using Multi-wavelength Analysis of Galaxy Physical Properties (MAGPHYS; \citealt{DaCunha2008}), a software package that fits galaxy SEDs using physical parameters of galaxies at the same redshifts and in the same photometric bands. 

MAGPHYS fits SEDs from far-UV to far-IR, so we use photometry from that range for fitting OHM host SEDs. In total, we use photometry from eight sources: the \textit{Galaxy Evolution Explorer} (\textit{GALEX}; \citealt{Martin2005}), the Sloan Digital Sky Survey (SDSS; \citealt{Stoughton2002}), the Two Micron All-Sky Survey (2MASS; \citealt{Skrutskie2006}), the \textit{Wide-field Infrared Survey Explorer} (\textit{WISE}; \citealt{Wright2010}), the Infrared Array Camera (IRAC) and Multiband Infrared Photometer for \textit{Spitzer} (MIPS) (both on \textit{Spitzer}; \citealt{Werner2005}), the \textit{Infrared Space Observatory} (\textit{ISO}; \citealt{Kessler1996}), and the \textit{Infrared Astronomical Satellite} (\textit{IRAS}; \citealt{Beichman1988}). In total, from these sources, we use up to 33 bands to fit OHM SEDs. We omit \textit{WISE} band 1 (3.4 μm) if \textit{Spitzer} IRAC band 1 (3.6 μm) exists for a given galaxy because the introduction of both causes poor fits, and IRAC tends to have smaller uncertainties than \textit{WISE}. The same is done for \textit{WISE} and IRAC band 2 (4.6 and 4.5 μm, respectively).

\begin{figure*}[t]
\centering
\includegraphics[width =0.49\textwidth]{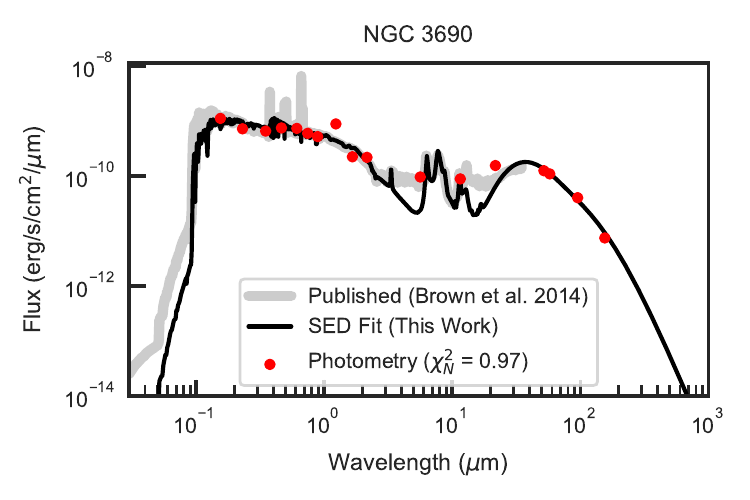}
\includegraphics[width =0.49\textwidth]{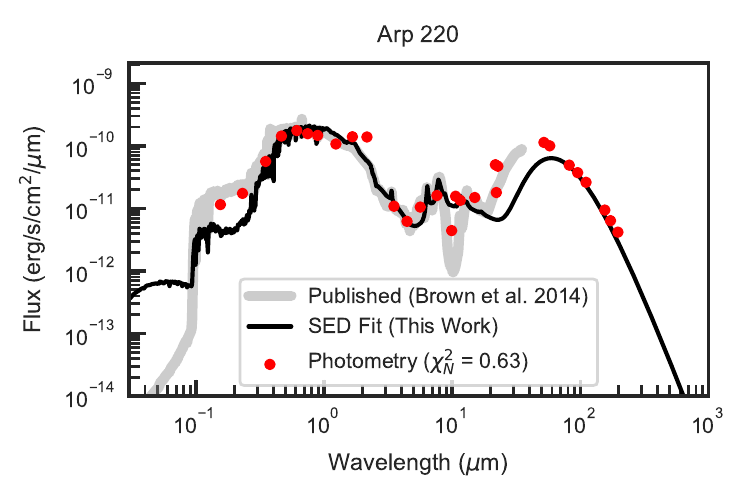} 
\caption{\label{fig:SED_exs} Examples of SED fits with MAGPHYS (black line). The published SED comes from \cite{Brown2014} and is shown as a thick grey line. Red dots denote the photometry used in a given fit.}
\end{figure*}

In Figure \ref{fig:SED_exs}, we present examples of these SED fits. For comparison, we use the Atlas of Galaxy SEDs \citep{Brown2014}, which includes a total of seven OHM host galaxy SEDs. The fits we present are imperfect matches to the complete \cite{Brown2014} SEDs, and we emphasize that our SED fits were done with the limited scope of reproducing observations of OHMs in our particular wavelength regime of interest, UV to mid-IR. Our fits are limited in wavelength outside this range, particularly in the far-IR. Although far-IR photometry would provide very useful information, data are sparse and unavailable for our objects of interest, so we have chosen to omit far-IR photometry. 
\newpage
\subsubsection{Emulating OHM Host Photometry} \label{sec:making_OH_data}
We use our SED fits to model OHM host galaxy observations for various missions and surveys as a function of redshift. We use PYPHOT,\footnote{PYPHOT's documentation can be accessed at \url{https://mfouesneau.github.io/docs/pyphot/}.} a package for calculating an object's photometry from its SED. It calculates the photometry using a given filter's transmission curve, $T(\lambda)$, by calculating the photon number flux:
\begin{equation}
N_\mathrm{tot} = \frac{1}{hc} \int_\lambda f_\lambda\ \lambda\ T(\lambda)\ d\lambda,
\end{equation}
where $f_\lambda$ is the flux density as shown above in Figure \ref{fig:SED_exs}. 

The greatest benefit of having SED fits is the ability to redshift them and mimic observations at higher redshifts. This scaling is done by adjusting the rest wavelength and the flux density by the inverse square of the luminosity distance and a redshift factor ($\propto D_L^{-2}\ (1+z)^{-1}$) and then re-``observing'' the OHM host. This method is used to create synthetic observations out to the desired redshifts.
\subsubsection{Emulating \HIit Host Galaxy Photometry}
For consistency, \HI host galaxy photometry is created similarly. However, instead of fitting SEDs, we use 57 SEDs\footnote{This number was originally 58 SEDs; however, after some examination, NGC 7674 seems to behave much more like an OHM in the near- to mid-IR like just a spiral galaxy, despite looking like a classic spiral morphologically. We attempted to do follow-up observations to determine if it potentially possessed both emission lines; however the OH line cannot be observed due to RFI. We therefore removed this galaxy from our \HI SED sample.} published in the Atlas of Galaxy SEDs \citep{Brown2014} that have previous \HI detections, a population of mainly spiral galaxies. Since these sources are not drawn from a strictly \HIns  -selected sample, they may not behave identically to samples from untargeted \HI surveys, although we expect differences to be modest. These SEDs are also redshifted, and photometry is ``measured'' using PYPHOT. 

\subsection{Machine Learning to Distinguish OH from \HIit}
To aid in determining if an emission line is an \HI or OH detection, we use machine learning algorithms to determine the likelihood of the line's classification. We employ a $k$-Nearest Neighbors ($k$-NN) algorithm that classifies objects based on a plurality vote of their neighbors' classes, where neighbors are determined within some parameter space \citep{Goldberger2005}. $k$-NN classification is a non-parametric method and a \textit{lazy learning} algorithm. Lazy learning means that the algorithm itself does not make assumptions or generalizations based on the training data, but instead uses those data to make direct decisions about the testing data. Algorithm ``optimization'' is purely done by our choices about the nearest-neighbor algorithm parameters used. The cost of using a lazy learning algorithm is the computation time in the testing phase. However, we are not testing on data sets large enough for slow speed to be problematic. This context makes the $k$-NN classification a robust and transparent method for our purposes. 

The final classification parameters are \textit{WISE} magnitudes and colors as well as the observed line frequency.  \cite{Suess2016} demonstrate that \textit{WISE} photometry can separate OHM and \HI populations at low redshift. We also choose \textit{WISE} because of its all-sky coverage, allowing it to be applicable to many different \HI surveys. Section \ref{sec:WISE_ML} discusses the use of \textit{WISE} magnitudes and colors to distinguish \HI and OH populations, as well as the limitations of using \textit{WISE} and its similarities to IRAC. Section \ref{sec:IRAC_ML} presents similar exercises using IRAC data, which have significant coverage over the LADUMA field but are otherwise less broadly applicable for other \HI surveys.

\subsubsection{OH and \HIit Classification Using WISE} \label{sec:WISE_ML}
The analysis in \cite{Suess2016} is done with low-redshift ($z<0.1$) objects and uses \textit{WISE} bands W1, W2, W3, and W4 (3.4, 4.6, 12, 22 μm).  W3 and W4 are very insensitive compared to W1 and W2; thus, this method is limited by both object brightness and redshift. We focus on using machine learning to sort using only W1 and W2 magnitudes, W1$-$W2 color, and the observed line frequency. 

\begin{figure}[t]
\centering
\includegraphics[width =\colw]{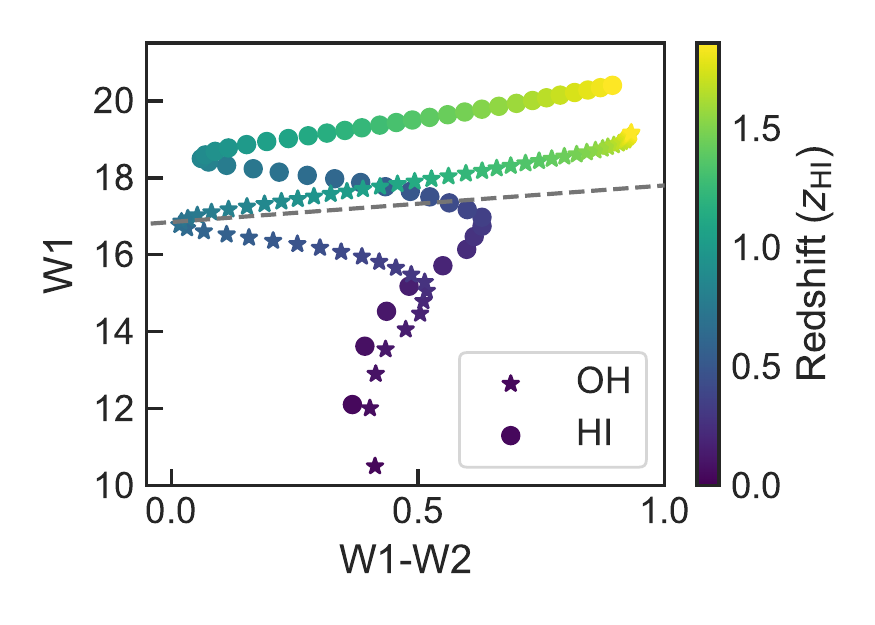}
\caption{\label{fig:W1vsW1W2_evolution} Predicted \textit{WISE} magnitude versus color for an OHM host galaxy (stars) and an \HI galaxy (circles). \HI redshift is denoted by color with redshift increasing with lightness. The corresponding OH points lie at the same observed frequency but are at a higher actual redshift. The grey dashed line represents the detection limit of \textit{WISE} (the region above the line is undetectable by \textit{WISE} band W1).}
\end{figure}

One of the cuts from \cite{Suess2016} is done in the color-magnitude space of W1 versus W1$-$W2 (or [3.4] versus [3.4]$-$[4.6]). We use this same parameter space for the $k$-NN algorithm. Examples of redshift evolution in this space are shown in Figure \ref{fig:W1vsW1W2_evolution} for an OHM host and an \HI source. These data are ``measured'' from their SEDs and show how OH-\HI separability varies with redshift.

We use the 57 \HI and 111 OHM host SEDs to test and train the $k$-NN algorithm. Each SED is redshifted to the maximum redshift detectable by LADUMA ($z_\mathrm{HI} = 1.45$ and $z_\mathrm{OH} = 1.876$), with \textit{WISE} photometry being ``measured'' roughly every $dz = 0.01$. We then remove any data too faint to have a $5\sigma$ WISE detection.

$k$-NN algorithms require feature scaling or parameter normalization, since the algorithm is inherently built on the distances between a data point and its neighbors. Therefore, we normalize each parameter from 0 to 1. $k$-NN classification algorithms are dependent on a few parameters that can be optimized for a given case. Parameters that were varied and tested for our purposes include the number of neighbors that is included in the plurality vote on an object's classification ($k$), whether neighbors are weighted by their distances, and how distance between objects is calculated ($p$). The distance between points is defined by the Minkowski distance of order $p$:

\begin{equation}
D(X,Y)=\Big(\sum_{i=1}^n |x_i-y_i|^p\Big)^\frac{1}{p},
\end{equation}
where $X$ and $Y$ are two points in an $n$-dimensional parameter space. Euclidean distance is recovered for $p=2$.
 
All algorithm optimization for this work is done by maximizing OH \textit{recall}. In machine learning classification, two metrics that are often considered when optimizing are \textit{precision} and \textit{recall}, both of which scale from 0 to 1 (1 being the best score). \textit{Precision} is the fraction of positive identifications that are correct. By optimizing precision, the number of false positives (or Type I errors) is minimized. \textit{Recall}, conversely, is the fraction of positives that were correctly identified. When recall is optimized, the number of false negatives (or Type II errors) is minimized \citep{Sammut2011}. These terms correspond to the familiar astronomical concepts of sample purity and completeness. In our case, a positive identification is the classification of a galaxy as an OHM host. We choose to optimize OH recall due to the rarity of OHMs and the desire to not miss any potential candidates. Although this approach increases the number of false positives, this algorithm does add information, and any positive identification it makes can motivate follow-up observations for confirmation. 

\begin{figure*}[t]
\centering
\includegraphics[width =\textwidth]{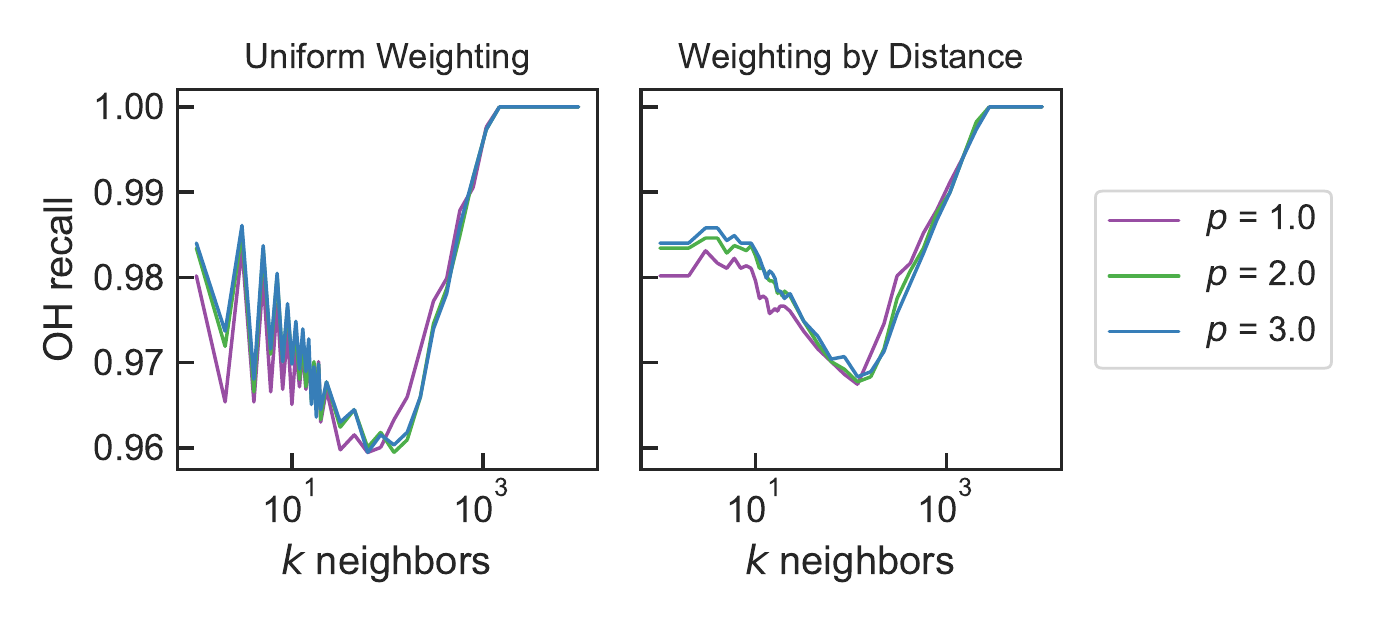}
\caption{\label{fig:kNN_optimization} Parameter exploration for our $k$-Nearest Neighbors ($k$-NN) classification algorithm. We vary the number of neighbors considered in voting ($x$-axis), whether distance is weighted (left and right plots), and how distance is determined using the Minkowski metric (the lines plotted). Each test is done using five-fold cross-validation. }
\end{figure*}

The algorithm parameter exploration is shown in Figure \ref{fig:kNN_optimization}. The $x$-axis shows a wide range of choices for the number of neighbors used, the lines plotted show a few choices for Minkowski distance, and the two panels show the difference between weighting and not weighting neighbors by distance. Each unique combination of parameters is tested using a five-fold cross-validation test. This process involves randomly sorting our data into two sets: training and testing data. The training data build the algorithm and the testing data determine how successful the algorithm is. This split was done five times, randomly splitting data each time, for each combination of parameters, and the final OH recall was determined by averaging the five individual OH recall values. In total, 2,000 $k$-NN algorithms were tested.

The results of these parameter tests give the highest OH recall for large numbers of neighbors ($k>10^3$). Although an OH recall of 1 would be ideal, this result comes at the cost of very low precision and defeats the purpose of the $k$-NN method by classifying based on the value of an algorithmic parameter instead of position relative to neighbors. Large numbers of neighbors also make for very computationally expensive algorithms. One common approach is to select $k=\sqrt{N}$, where $N$ is the number of data points, but for our data ($k\approx140$), this choice of $k$ is near the lowest value of OH recall. Another common approach is to select $k=1$ or another low number. For small data sets, always assuming the nearest object has the same classification can introduce noise. However, with a sufficiently large data set, this trend is less problematic. Figure \ref{fig:kNN_optimization} indicates that a small $k$ achieves a recall of over 0.98 for distance-weighted learning. We stress that there exists no optimal $k$ for all purposes, since each $k$-NN optimization varies based on the properties of the data \citep{Altman1992a}.

We choose our number of neighbors to be $k=3$, based on the above considerations. Increasing a small amount above $k=1$ also reduces noise while maximizing OH recall. For weighting and Minkowski metric, we choose to weigh votes by distance and use standard Euclidean distance ($p=2$).

\begin{figure}[t]
\centering
\includegraphics[width =\colw]{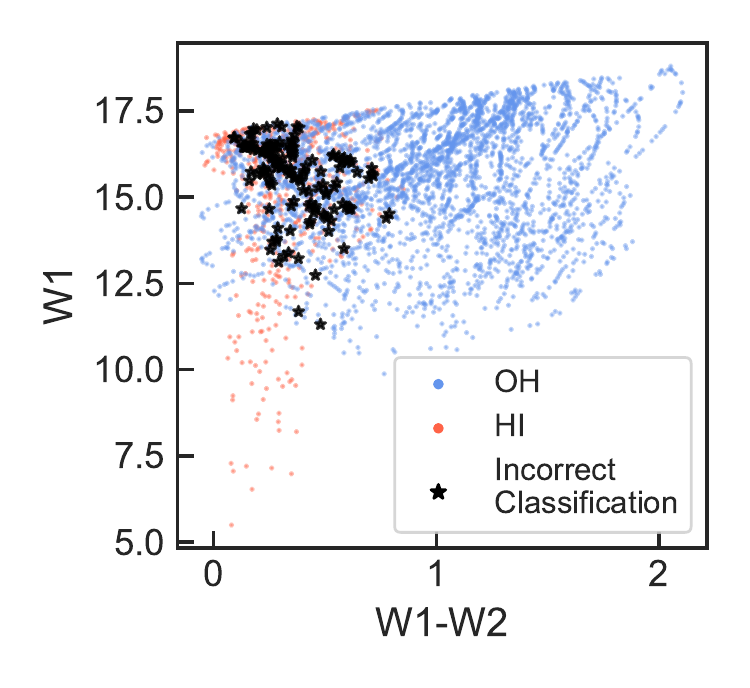}
\includegraphics[width =\colw]{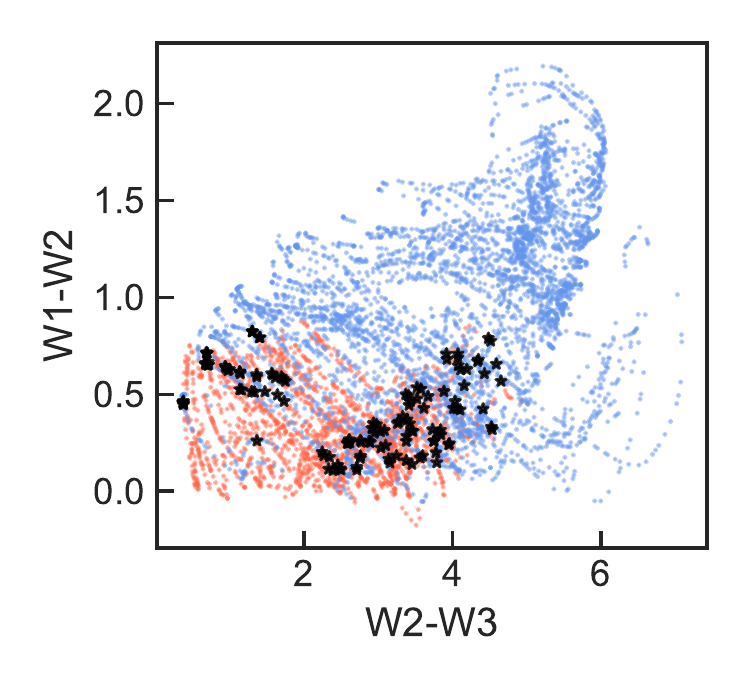}
\caption{\label{fig:WISE_ML} Final results from training and testing our $k$-NN algorithm using \textit{WISE} W1 versus W1$-$W2 (above) and W1$-$W2 versus W2$-$W3 (below). Blue points indicate OH host testing points that were correctly identified, and red indicate the same for \HI sources. Black stars show the misidentified objects, either OH misidentified as \HI or vice versa (3.6\% of objects in top panel, 1.9\% of objects in bottom panel).  Note that the OH and \HI markers are partially transparent to show overlapping.}
\end{figure}

The final trained and tested $k$-NN algorithm results are shown in the top panel of Figure \ref{fig:WISE_ML}. The blue and red points show correctly identified OH and \HI host galaxies respectively, and the black stars are misidentified objects. The incorrect identifications concentrate where OH and \HI sources overlap the most at the same observed frequency, and indicate where there will be the most confusion. The final OH recall is \textbf{0.985} and the OH precision is \textbf{0.974}. In other words, for redshifts less than $z\sim 1.0$, we expect to identify 98.5\% of OH lines in \HI surveys, thereby mitigating the impact of contamination.

We repeat this process for another \cite{Suess2016} parameter space, W1$-$W2 versus W2$-$W3, as well as the observed line frequency. This alternative approach significantly limits the number of available detections because of the inclusion of the comparatively less sensitive W3 band. This approach however leads to a higher OH precision, since mid-IR data are relevant for distinguishing between these populations. The results of this test are shown in the bottom panel of Figure \ref{fig:WISE_ML}. Precision and recall from this test are compared to those for other tests in Table \ref{tab:ML_results}. 

\subsubsection[t]{OH and \HIit Classification Using IRAC} \label{sec:IRAC_ML}
As discussed previously, \textit{WISE} is insensitive to galaxies at redshifts above $z\sim 1.0$. Although having significantly less sky coverage than \textit{WISE}, \textit{Spitzer}'s IRAC bands 1 and 2 are very similar to \textit{WISE} bands 1 and 2 but are much more sensitive and can detect OHM and \HI host galaxies over the full redshift ranges probed by both LADUMA and SKA1. We therefore perform an exercise similar to that in Section \ref{sec:WISE_ML} using IRAC data. Throughout this paper, IRAC bands are referred to by their wavelengths in microns (e.g., IRAC [3.6] denotes the 3.6 $\mu$m band magnitude).

\begin{figure}[t]
\centering
\includegraphics[width =\colw]{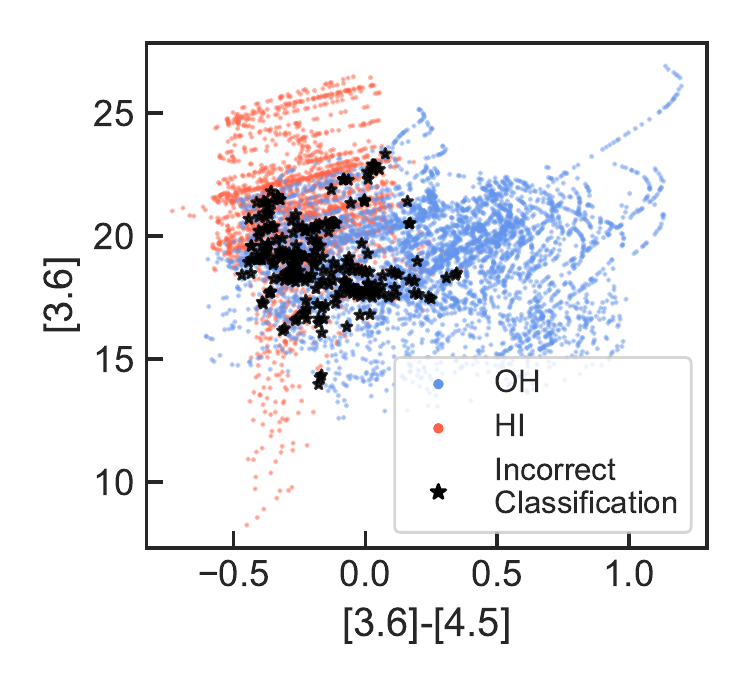}
\includegraphics[width =\colw]{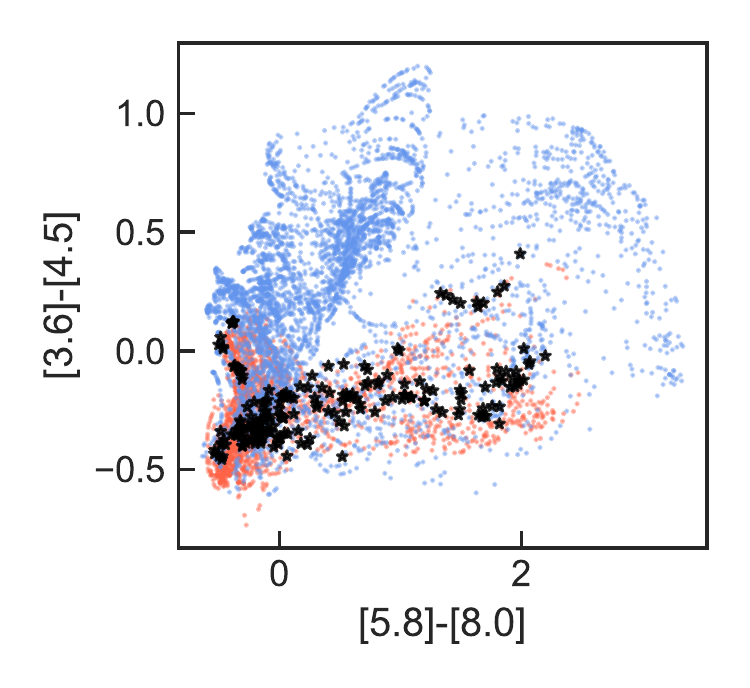}
\caption{\label{fig:IRAC_ML} Final results from training and testing our $k$-NN algorithm using \textit{Spitzer} IRAC [3.6] versus [3.6]$-$[4.5] (above) and [3.6]$-$[4.5] versus [5.8]$-$[8.0] (below). Blue points indicate OH host testing points that were correctly identified, and red indicate the same for \HI sources. Black stars show misidentified objects (3.4\% of objects in top panel, 2.9\% of objects in bottom panel). Note that the OH and \HI markers are partially transparent to show overlapping.}
\end{figure}

We use a parameter space analogous to the first test for the IRAC $k$-NN algorithm ([3.6] versus [3.6]$-$[4.5]). (As a reminder, \textit{WISE} uses Vega-based magnitudes, whereas IRAC uses AB magnitudes.) Since IRAC is sensitive to the entire redshift range of our \HI and OHM hosts, we do not perform any detection cuts. Results from this exercise are presented in Table \ref{tab:ML_results} and visualized in Figure \ref{fig:IRAC_ML}. Comparing to the analogous \textit{WISE} space, this test has the same OH precision, but OH recall suffers slightly. However, achieving an OH recall of 0.979 is still a powerful tool when it comes to sorting OHM hosts from \HI hosts, and the ability to probe to higher redshifts has strong appeal.

We also consider IRAC [3.6]$-$[4.5] versus [5.8]$-$[8.0]. \cite{Stern2005} suggest that cuts in this space can separate active galaxies from normal galaxies. We perform algorithm optimizations similar to those mentioned previously before training and testing. We present the final results in Table \ref{tab:ML_results} and Figure \ref{fig:IRAC_ML}. Despite having information from the [8.0] band, this test performs slightly worse than the previous tests in both OH recall and precision, indicating that the overlap between OHM and \HI hosts is greater in this parameter space than in the previous alternatives. 

Being able to probe the full redshift range of LADUMA is beneficial, but inevitably introduces more contamination, as indicated by the slightly reduced OH recall. However, it is worth noting that despite being lower, these recall values still exceed 95\%. These IRAC and \textit{WISE} tests create a new framework for the process of separating OH and \HI host populations.

\begin{deluxetable}{cccc}
\tablecaption{Machine Learning Results for Distinguishing \HI Emission Lines from OH Megamasers\label{tab:ML_results}}\tablehead{\colhead{Mission} & \colhead{Features} & \colhead{OH Recall} & \colhead{OH Precision}} 
\startdata
\textit{WISE} & W1, W1$-$W2, $\nu$ & 0.985 & 0.974 \\
\textit{WISE} & W1$-$W2, W2$-$W3, $\nu$ & 0.987 & 0.985 \\
IRAC & [3.6], [3.6]$-$[4.5], $\nu$ & 0.979 & 0.972 \\
IRAC & [3.6]$-$[4.5], [5.8]$-$[8.0], $\nu$ & 0.982 & 0.976 \\
\enddata
\tablecomments{The features column indicates what data were used to distinguish \HI  from OHM host galaxies. Each row includes the observed frequency ($\nu$) of the line in question (OH or \HIns) to assist in separation of sources. }
\end{deluxetable}

\section{Discussion} \label{sec:discussion}
The methods presented in this paper will be crucial to mitigating OHM contamination of \HI emission-line surveys. \textit{WISE} provides the all-sky coverage needed for upcoming surveys that will be covering large portions of the sky, while IRAC has the deep-field coverage needed for surveys such as LADUMA, which will be the deepest \HI emission-line survey to date.  

One of the biggest shortcomings of these methods and calculations is that they are based on known OHMs, which currently extend to a highest redshift of $z_\mathrm{OH}=0.264$ \citep{Darling2002a}. This limitation has forced us to make some extrapolations to obtain predictions for higher-redshift surveys. This approach is unavoidable until we have higher-redshift data on both \HI and OH populations. As more surveys are conducted, we will be able to update the OHLF and OH SED evolution as well as provide tighter constraints on these calculations and predictions. 

As discussed in Section \ref{sec:intro}, \HI and OH sources can be separated by spectroscopic redshift. It is therefore worth recognizing that some objects will be readily identifiable and that these objects will be crucial for helping classify those without redshifts. For LADUMA, the current largest source of spectroscopic redshifts is the PRIsm MUlti-Object Survey (PRIMUS), with over 32,000 redshifts in the field and in the relevant redshift range \citep{Coil2010}. PRIMUS only detects galaxies out to $z\sim 1.2$, meaning that some of the most potentially contaminated (i.e., highest) redshift ranges will have few spectroscopic redshifts available. Another source of spectroscopic redshifts soon to come online is the Wide-Area VISTA Extragalactic Survey (WAVES), which will have two campaigns, WAVES-Wide (large-sky, low-redshift) and WAVES-Deep (small-area, high-redshift) \citep{Driver2019}. WAVES-Deep will have several small patches, including one on the LADUMA field. Slated to target 45,000 objects, WAVES-Deep will also be crucial in identifying objects; however the current estimates only show detections out to redshift $z\sim 0.8$. For future all-sky untargeted \HI surveys such as those on ASKAP or the SKA, WAVES-Wide aims to provide 880,000 spectroscopic redshifts out to redshift $z\sim 0.2$. Although we may have many redshifts for identifying objects as OH or \HI sources, these redshifts are extremely limited where potential OH contamination is the greatest threat.

\begin{figure*}[t]
\centering
\includegraphics[width=\textwidth]{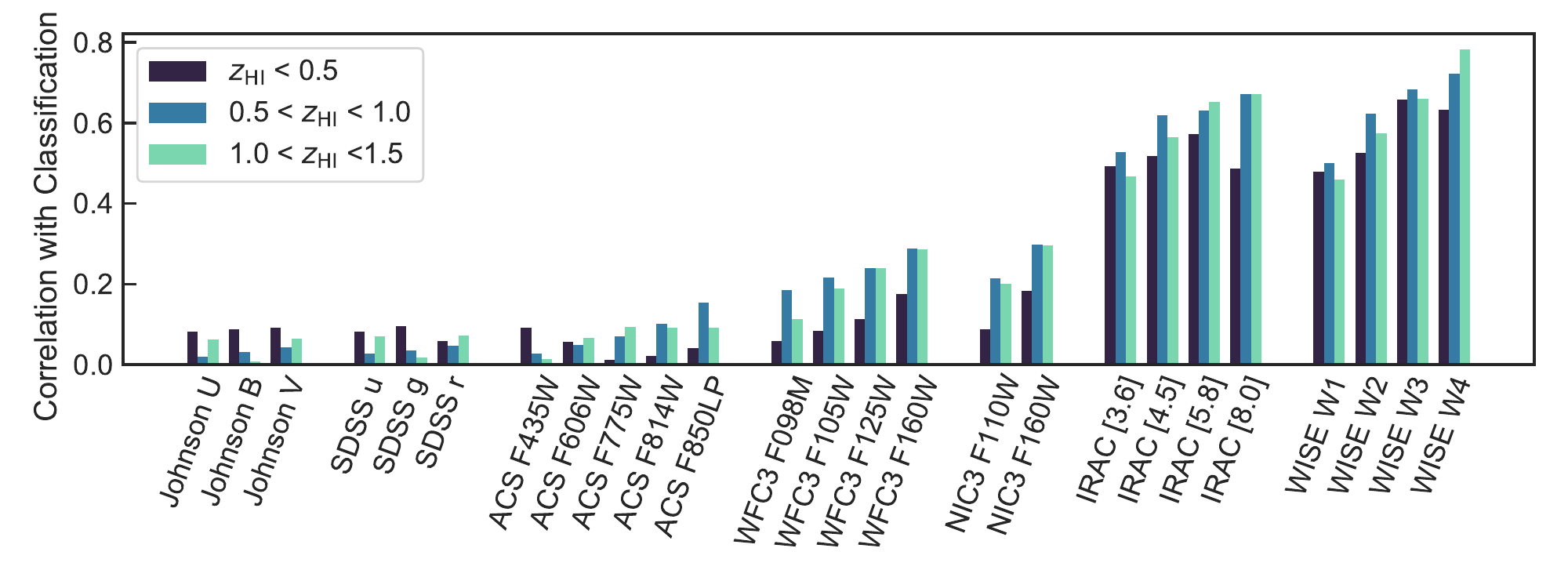}
\caption{\label{fig:correlation_vs_band} Correlation of band photometry with OH/\HI classification using a Pearson correlation test. Bands on the $x$-axis are grouped by mission or type and then ordered by increasing wavelength. Each correlation is tested in three redshift ranges.}
\end{figure*}

\textit{WISE} and IRAC photometry were not the only data tested in Section \ref{sec:separating} for the ability to separate OHM and \HI hosts. We also tested other photometry for separability, focusing on data that have significant coverage in the LADUMA field. These include SDSS \textit{ugr}, Johnson \textit{UBV}, and \textit{HST} ACS, WFC3, and NICMOS bands. Figure \ref{fig:correlation_vs_band} shows how each of these bands correlates with OH/\HI classification. For each band, a Pearson correlation test was done for three redshift ranges. Bands are grouped on the $x$-axis and then sorted by increasing wavelength. Figure \ref{fig:correlation_vs_band} demonstrates the sorting value of bands in the near- to mid-IR. This distinction is due to the extreme star formation in OHM host galaxies, which is detected in the IR. Optical bands are poorer candidates for separation, since they are less sensitive to high star-formation rates in dusty, gas-rich systems.

\section{Conclusions} \label{sec:conclusions} 
We present predictions for the numbers of OH megamasers that will be detected in future untargeted \HI surveys and explore how those numbers impact \HI source confusion over a range of redshifts up to $z_\mathrm{HI}=1.5$. To assist in untangling these populations, we also present methods for estimating the likelihood that a line has been identified as \HI or OH. Below, we summarize our predictions and discuss the implications of this work for future \HI surveys: 
\begin{enumerate}
\item LADUMA will likely triple the number of known OHMs: we predict $83^{+21}_{-17}$ new detections. Larger surveys with telescopes such as the SKA1 will detect thousands more OHMs.

\item The contamination these OHM detections will impose on \HI line surveys is highly dependent on redshift (and, secondarily, depth). In a line flux-limited survey, OHMs are more abundant at higher redshift, while \HI sources become sparser. For these high-redshift surveys, OH detections will outnumber \HI detections near redshift $z_\mathrm{HI} \sim 1.0$. 

\item Near- and mid-IR observations can assist in separating \HI from OHM emission lines, which we demonstrate using a $k$-Nearest Neighbors machine learning algorithm. We will be able to identify nearly 99\% of OH lines for redshifts less than $z\sim 1.0$ and 96\% of lines at higher redshifts.
\end{enumerate} 

Although OHM host galaxies represent a potential contamination for untargeted \HI line surveys, these rare and interesting objects can be important scientific tools. As discussed in Section \ref{sec:mergerrate}, OHM density can also provide an independent measurement on the major merger rate evolution paramater, $\gamma$, since OHMs serve as tracers of major galaxy mergers. These galaxies are signposts of the most extreme star formation in our universe, signaling where the most massive starbursts are happening \citep{Briggs1997}, and can even offer a way to measure in-situ magnetic fields using Zeeman splitting \citep{Robishaw2008,McBride2014}. The methods presented in this paper and follow-up observations will begin uncovering these galaxies and allowing us to characterize them at higher redshifts and potentially create better methods for mitigating contamination in \HI surveys. 

\acknowledgements
This work has been supported by the National Science Foundation through grants AST-1814648 (to HR and JD) and AST-1814421 (to AB). This research has made use of the NASA/IPAC Extragalactic Database (NED), which is funded by the National Aeronautics and Space Administration and operated by the California Institute of Technology. We thank Sarah Blyth, Natasha Maddox, and Aaron Stemo for helpful insights and conversations. We also thank the anonymous referee for their thorough and helpful comments which improved the presentation of this paper.
\newline
\software{Astropy \citep{astropy}, Matplotlib \citep{matplotlib}, Numpy \citep{numpy}, Scipy \citep{scipy}, Scikit-Learn \citep{sklearn}, MAGPHYS \citep{DaCunha2008}, emcee \citep{emcee}, PYPHOT (\url{https://github.com/mfouesneau/pyphot})}

\bibliography{library}
\end{document}